# Tunable electronic structure and magnetic anisotropy in bilayer ferromagnetic semiconductor $Cr_2Ge_2Te_6$

Wen-ning Ren[1,2], Kui-juan Jin[1,2,3,*], Jie-su Wang[1], Chen Ge[1,2], Er-Jia Guo[1,2], Cheng Ma[1,2], Can Wang[1,2,3] & Xiulai Xu[1,2,3]

[1]*Beijing National Laboratory for Condensed Matter Physics, Institute of Physics, Chinese Academy of Sciences, Beijing 100190, China*

[2]*School of Physical Sciences, University of Chinese Academy of Sciences, Beijing 100049, China*

[3]*Songshan Lake Materials Laboratory, Dongguan 523808, China*

[*]Correspondence and requests for materials should be addressed to Kuijuan Jin:

kjjin@iphy.ac.cn

**The emergence of ferromagnetism in two-dimensional van der Waals materials has aroused broad interest. However, the ferromagnetic instability has been a problem remained. In this work, by using the first-principles calculations, we identified the critical ranges of strain and doping for the bilayer $Cr_2Ge_2Te_6$ within which the ferromagnetic stability can be enhanced. Beyond the critical range, the tensile strain can induce the phase transition from the ferromagnetic to the antiferromagnetic, and the direction of magnetic easy axis can be converted from out-of-plane to in-plane due to the increase of compressive strain, or electrostatic doping. We also predicted an electron doping range, within which the ferromagnetism can be enhanced, while the ferromagnetic stability was maintained. Moreover, we found that the compressive strain can reverse the spin polarization of electrons at the conduction band minimum, so that two categories of half-metal can be induced by controlling electrostatic doping in the bilayer $Cr_2Ge_2Te_6$. These results should shed a light on achieving ferromagnetic stability for low-dimensional materials.**

Since 2004 when the graphene was exfoliated by Geim and Novoselov [1], researchers have revealed many unique physical properties from various two-dimensional (2D) materials, e.g. quantum spin Hall candidate monolayer $WTe_2$ [2], stanine [3] with topological band inversion, and high-mobility black phosphorus [4,5]. Nevertheless, the absence of intrinsic ferromagnetism limits their application in spintronic devices. Recently, the intriguing intrinsic ferromagnetism has been proved both theoretically [6-9] and experimentally [10] in $Cr_2Ge_2Te_6$, one of the layered transition metal trichalcogenide's family, which broke the long-established Mermin-Wagner theorem [11] and greatly enriched the versatility of 2D materials. Some applications have been proposed in new-generation magnetic memory storage devices [12] and nanoelectronic devices [13].

As the existence of ferromagnetism is one of the most charming features in 2D layered materials, it is important to enhance the stability and realize the tunability of the long-range magnetic ground state. An effective avenue is to increase the magnetic anisotropy energy (MAE), which is based on the energy difference between the in-plane and out-of-plane magnetization direction. Large MAE in van der Waals (vdW) magnets would lift Mermin-Wagner restriction [11,14], for that the out-of-plane magnetic anisotropy would open a spin-wave gap and counteract magnetic fluctuations, resulting in the stabilization of the long-range ferromagnetic order [9,15,16]. The modulation of the magnetic properties based on the band engineering is highly desired in 2D layered ferromagnets, and the applications of external electric field [17-21], pressure [22,23], electrostatic doping [24-26], and strain engineering [7,27-31] offer some valid approaches to tuning electronic structures, as well as the physical properties. Although revealing the transformation of MAE under external factors is imperative, so far, few researches has focused on the tunability of MAE with strain engineering or electrostatic doping for bilayer $Cr_2Ge_2Te_6$.

In this letter, the first-principles calculations were carried out to study the tunability of electronic structures and the magnetism in bilayer $Cr_2Ge_2Te_6$ with biaxial strain or electrostatic doping. We determined a critical range of strain or doping in which the MAE is increased, in other words, the ferromagnetic stability is enhanced. A range of electron doping is also predicted, within which the ferromagnetism and the (Curie temperature) $T_c$ can be raised. We also showed that two types of half-metal were induced based on the external regulation. We further explored the possible mechanism involved in the variations of MAE, which would provide deeper understanding of 2D ferromagnetic materials.

## Results and discussion

The crystal structures from top and side views of bilayer $Cr_2Ge_2Te_6$ are shown in Fig. 1a and b, respectively. The unit cell of bilayer $Cr_2Ge_2Te_6$ is denoted by the gray shaded region, which contains four Cr atoms, and each of them is bonded to six nearest-neighboring Te anions and locates at the center of an octahedron (denoted by light green) formed by these Te atoms. Each layer consists of a honeycomb network of Cr atoms similar to graphene and comprises a Ge-Ge metal bond [32], which is a dimer lying perpendicularly at the central position of the $CrTe_6$ nets, forming ethane-like groups of $Ge_2Te_6$. The bilayer $Cr_2Ge_2Te_6$ contains two layers placed in AB stacking sequence. According to the structure optimization, the vdW gap in the bilayer is 3.437 Å, and the determined lattice parameters is $a = b = 6.838(2)$ Å, as listed in Table 1.

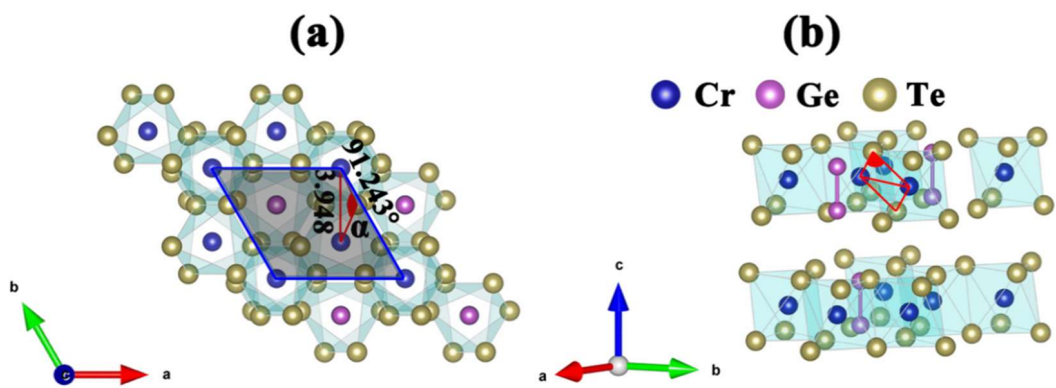

**Figure 1.** (**a**) Top and (**b**) side views for pristine bilayer $Cr_2Ge_2Te_6$ in AB stacking. Blue, purple, and yellow balls represent Cr, Ge, and Te atoms, respectively. The gray shaded region denotes the unit cell. The red shaded areas indicate the Cr-Te-Cr angle $\alpha$ and the bond between the nearest neighbor Cr atoms are connected by red solid lines.

We firstly explore the lattice distortion with the biaxial strain and electrostatic doping. Here, the strain is denoted by $\eta = (a/a_0 - 1) \times 100\%$, where $a$ and $a_0$ correspond to the strained and pristine lattice constants (without any strain or doping), respectively. So, the positive sign of $\eta$ represents tensile strain, and the negative sign of it represents compressive strain. The electrostatic doping concentration is tuned by altering the total number of electrons in a unit cell. The positive and negative signs of concentration represent hole and electron doping, respectively. Figures 2a and b exhibit how the Cr-Te-Cr angles $\alpha$ and Cr-Cr bond $d$ change versus biaxial strains. According to the previous theoretical predictions [7,8,22,33-35], since the Cr-Te-Cr angle is close to $90°$, the super-exchange interaction favors ferromagnetic (FM). The direct exchange interaction favors antiferromagnetic (AFM), which is inversely proportional to $d$ [36,37]. The competition between super-exchange interaction and direct exchange interaction can be effectively tuned by controlling Cr-Te-Cr angle and the nearest-neighbor Cr-Cr bond, which affects the magnetic ground state directly [22,34]. As shown in Fig. 2b, the $\alpha$ and $d$ decrease (increase) with the increase of compressive (tensile) strain. As shown in Fig. 2c, we define the lengths of bonds in Cr-Te as $l_a$ and $l_b$ of octahedra $a$ and $b$, respectively. $\bar{l}_a$ and $\bar{l}_b$ are the average bonds of six Cr-Te bonds in $a$ and $b$ octahedra, respectively. As shown in Fig. 2d, the $\bar{l}_a$ and $\bar{l}_b$ decrease (increase) with the increase of compressive (tensile) strain. The increase (decrease) in average bonds $\bar{l}_a$ and $\bar{l}_b$ means the octahedra expanding (shrinking) with the increase in the tensile (compressive) strain which is closely related to the magnetic and electronic structure, and the relations will be discussed below.

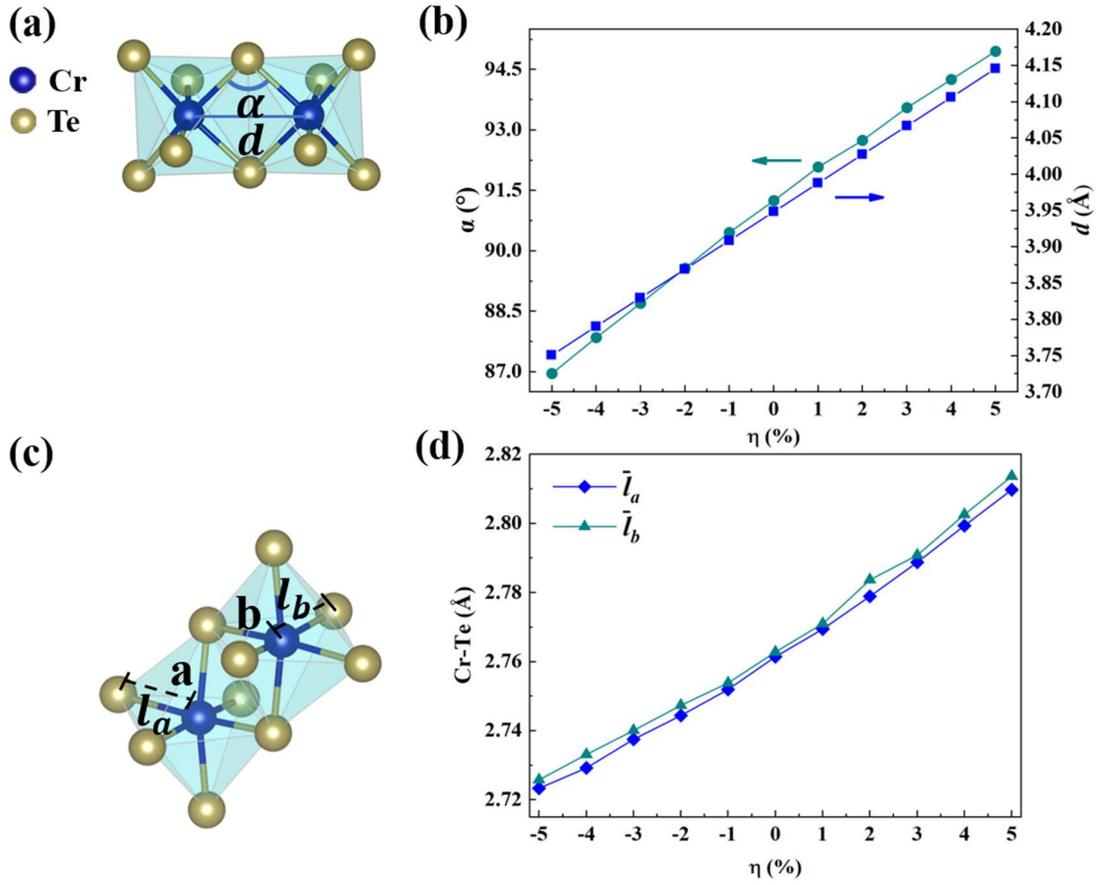

**Figure 2.** (**a**) Schematic illustration of Cr-Te-Cr angles $\alpha$ and Cr-Cr bond $d$. (**b**) Strain dependence of Cr-Te-Cr angles (left axis) and Cr-Cr bond lengths (right axis) for bilayer $Cr_2Ge_2Te_6$. The arrows point to the axises for each curve in the corresponding color. (**c**) The bond length of Cr octahedron a and b: $l_a$ and $l_b$, respectively. $\bar{l}_a$ and $\bar{l}_b$ denote the average bonds. (**d**) The average bonds $\bar{l}_a$ and $\bar{l}_b$ versus strain.

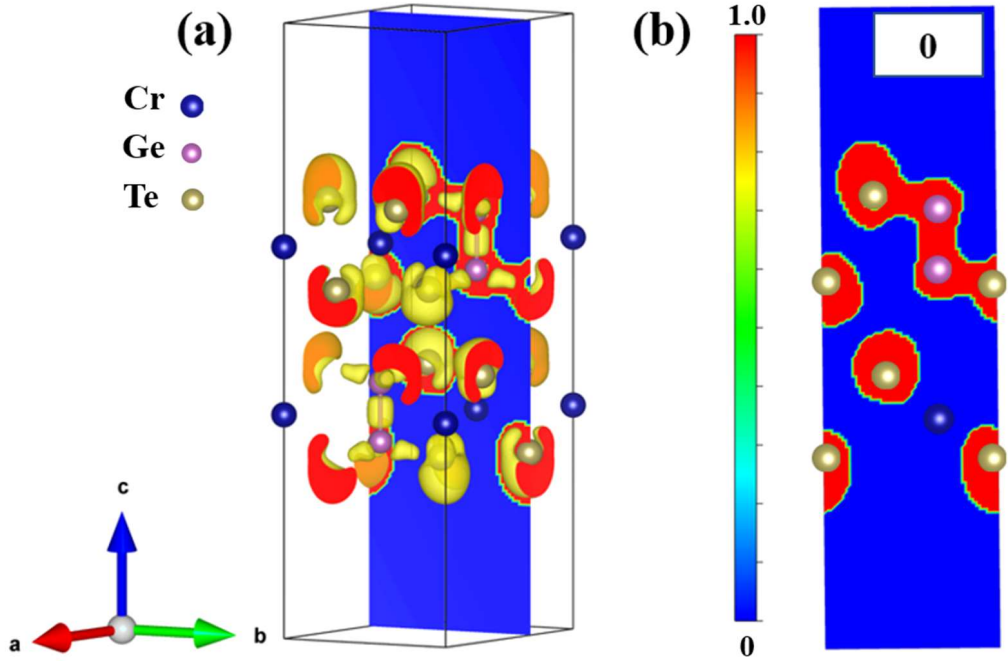

**Figure 3**. Electron localization function (ELF). (**a**) The isosurface demonstration of pristine (0) bilayer $Cr_2Ge_2Te_6$. (**b**) The two-dimensional contour map in the direction of [001], the cut surface is shown in blue.

To explore the electronic structures of the bilayer $Cr_2Ge_2Te_6$, the electron localization function (ELF) is simulated. Figure 3a shows the isosurface demonstration of pristine bilayer $Cr_2Ge_2Te_6$. In order to explain the electron localization distribution more clearly, the two-dimensional contour map in the direction of [001] is shown in Fig 3b, from which we can see the electron distribution around Te and Ge atoms is highly localized. The ELF value around Cr atom is 0, indicating the highly delocalization of electrons around Cr. The boundary between the localized and delocalized electron distribution is green, which means that the ELF value is about 0.5 and the ionic bond exists between Cr and Te atoms. We also studied the effects on the electron localization distribution around different kinds of atoms under biaxial strains or electrostatic doping concentrations, and found no obvious difference in the electron localization distribution.

The MAE is defined as $MAE = E_{[100]} - E_{[001]}$ for each unit cell (four Cr atoms), where $E_{[100]}$ and $E_{[001]}$ denote the total energy for the magnetic moments oriented along in-plane and out-of-plane, respectively. From this definition, we can determine that increased *MAE* means the enhanced ferromagnetic stability [14]. The detailed results are listed in Table 1, from which we can see the positive MAE for pristine bilayer $Cr_2Ge_2Te_6$, illustrating that the out-of-plane ([001]) direction is the easy axis for the magnetization. Also, we can see that the spin magnetic moment of Cr increases as the electron doping concentration increases from 0 to 0.2 e/u.c. Furthermore, from our results, the negative spin magnetic moment of Te relative to Cr is obtained, which is crucial for the stability of ferromagnetic ordering of Cr ions [38].

**Table 1.** The lattice constant $a$ and interlayer distance $l$ (vdW gap) of pristine (0) bilayer $Cr_2Ge_2Te_6$ used in the present calculations. $E_g$ and *MAE* stand for the band gap and magnetic anisotropy energy, respectively. Spin ($m_s^{Cr}$, $m_s^{Ge}$, $m_s^{Te}$) and orbital ($m_o^{Cr}$, $m_o^{Ge}$, $m_o^{Te}$) moments of the structures calculated by GGA+U with the spin-orbit coupling included. The results under specific compressive strain (-2%), tensile strain (1%), electron doping (-0.1 e/u.c.) and hole doping (0.1 e/u.c.) are also included.

| Structure | | $a$ (Å) | $l$ (Å) | $E_g$ (eV) | *MAE* (meV/u.c.) | $m_s^{Cr}$ ($m_o^{Cr}$) ($\mu_B$/at) | $m_s^{Ge}$ ($m_o^{Ge}$) ($\mu_B$/at) | $m_s^{Te}$ ($m_o^{Te}$) ($\mu_B$/at) |
|---|---|---|---|---|---|---|---|---|
| Bilayer | 0 | 6.838 | 3.437 | 0.374 | 0.162 | 3.233 (0.002) | 0.034 (0.001) | −0.115 (−0.002) |
| | -2% | 6.701 | 3.454 | 0.174 | 0.300 | 3.206 (0.006) | 0.027 (0.001) | −0.109 (−0.002) |
| | 1% | 6.907 | 3.369 | 0.446 | 0.222 | 3.248 (0.001) | 0.037 (0.001) | −0.118 (−0.002) |
| | −0.1 | 6.853 | 3.443 | – | 0.126 | 3.245 (0.002) | 0.035 (0.001) | −0.114 (−0.002) |
| | -0.2 | 6.867 | 3.447 | – | 0.062 | 3.255 (0.002) | 0.035 (0.001) | -0.114 (-0.002) |
| | 0.1 | 6.853 | 3.425 | – | 0.302 | 3.229 (0.002) | 0.033 (0.001) | −0.115 (-0.002) |

We then investigated the spin-polarized band structures of pristine bilayer $Cr_2Ge_2Te_6$ just for comparing, as shown in Fig. 4. From Fig. 4a, we can see that both conduction band minimum (CBM) and valence band maximum (VBM) are of purely spin-up character, possessing indirect band gap. The spin-polarized band structures of bilayer $Cr_2Ge_2Te_6$ with $\eta = -2\%$ are plotted in Fig. 4b. Compared that of the pristine one [Fig. 4a], the band gap is significantly reduced due to the increased bandwidth. Attractively, the spin-polarized character at the CBM has even changed from spin-up [Fig. 4a] to spin-down [Fig. 4b], indicating that the electrons at the CBM and VBM are with opposite spin. As shown in Fig. 4c, the larger compressive strain ($\eta = -5\%$) further increases the bandwidth due to the stronger interatomic coupling in the few-layer $Cr_2Ge_2Te_6$, inducing a semiconductor-metal phase transition. On the contrary, the tensile strain (1%) increases the band gap, and the spin polarization at the CBM is enhanced compared with the pristine $Cr_2Ge_2Te_6$, as shown in Fig. 4d. Interestingly, in the vicinity of the Fermi level, the compressive strain induces a degeneracy of the spin-down energy band along the K-M direction, which is marked in the dashed blue circle. The half-metallic state (conduction electrons being spin-up) is induced with electron doping (-0.1 e/u.c.) due to the obvious spin polarization, as shown in Fig. 4e. While with the hole doping (0.1 e/u.c.), the Fermi level shift down into the valence band, which makes the material metallic, as shown in Fig. 4f.

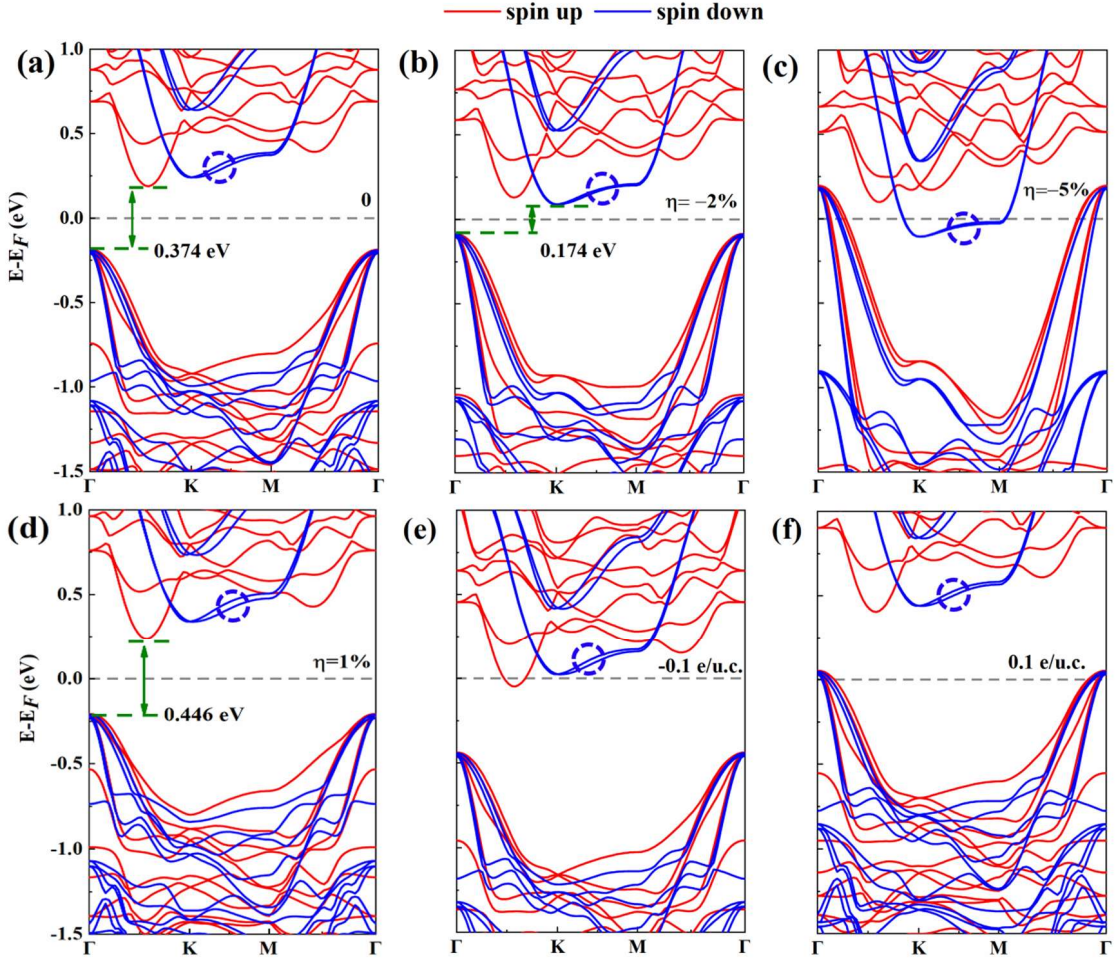

**Figure 4**. The spin-polarized electronic band structures of bilayer $Cr_2Ge_2Te_6$: (**a**) pristine, (**b**) compressive strain of 2%, (**c**) compressive strain of 5%, (**d**) tensile strain of 1%, (**e**) electron doping of 0.1 e/u.c, and (**f**) hole doping of 0.1 e/u.c. To achieve the visualizations of band gaps, the CBM and VBM are denoted by olive horizontal dashed lines, which are connected by double-headed olive arrows. The gray horizontal dashed lines denote the Fermi level. The degenerate spin-down energy band is marked in the dashed blue circle.

We further studied the band structure of bilayer $Cr_2Ge_2Te_6$ under dual regulations, demonstrating that the half-metallic state (conduction electrons being spin-down) can also be induced in bilayer $Cr_2Ge_2Te_6$ by combining electron doping (-0.1 e/u.c.) and compressive strain ($\eta = -2\%$), as shown in Fig. 5a. These results indicate the potential usage of few-layer $Cr_2Ge_2Te_6$ in spintronic devices. In Fig. 5b, the schematic of the electronic structures restructuring illustrates the regulation mechanism more visually.

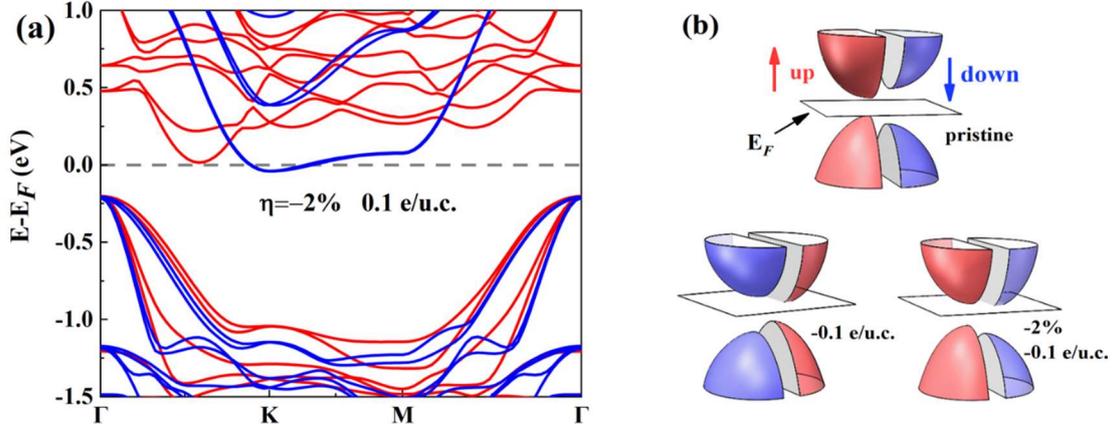

**Figure 5.** (**a**) The band structure of bilayer Cr$_2$Ge$_2$Te$_6$ under dual regulations (compressive strain of 2% and electron doping of 0.1 e/u.c.). Horizontal dashed lines denote the Fermi level. (**b**) The schematic of the electronic structures restructuring models under the electron doping (-0.1 e/u.c.) or dual regulations. Spin-up and spin-down states are marked by red and blue, respectively. The white plane denotes the Fermi surface.

Figure 6 shows the atomic projected density of states (PDOS) of the bilayer Cr$_2$Ge$_2$Te$_6$ under specific strains or electrostatic doping concentrations, as well as the schematic electron configurations for pristine or doping of -0.1 e/u.c. Here, only the PDOS of Cr $d$ orbitals are presented, which makes a major contribution to MAE based on the perturbation theory analysis. As shown in Fig. 6b, the spin polarization at the CBM of bilayer Cr$_2$Ge$_2$Te$_6$ is weakened under the strain of $\eta = -2\%$. Moreover, the states at CBM is transformed into spin-down, which originates mainly from Cr $d_{yz}/d_{xz}$ orbitals. As the compressive strain further increases ($\eta = -5\%$), the band gap disappears and the orbital overlaps. With tensile strain of $\eta = 1\%$, the contribution of Cr $d_{xy}/d_{x^2-y^2}$ orbitals near the Fermi level is reduced compared to the pristine one, as shown in Fig. 6d. As seen in Fig. 6e, the half-metallic state is induced due to the obvious spin polarization at the doping concentration of -0.1 e/u.c in the bilayer Cr$_2$Ge$_2$Te$_6$, and the spin-up states near the Fermi level are mainly from

Cr $d_{yz}/d_{xz}$ orbitals. To understand the increased magnetic moment for Cr atoms with electron doping (listed in Table 1), the diagrammatic electronic configurations are shown in Fig. 6g for the pristine or doped (-0.1 e/u.c.) bilayer $Cr_2Ge_2Te_6$. Compared with the pristine one, the Cr-$d_{yz}$ state is occupied by doped electrons, so that the net magnetic moment increases at the doping concentration of -0.1 e/u.c., indicating the enhanced ferromagnetism of the system. It is worth mentioning that Cr $d_{yz}$ and $d_{xz}$, as well as $d_{xy}$ and $d_{x^2-y^2}$ orbitals are degenerate because of the crystal symmetry. The shift of the Fermi level due to the strain or electrostatic doping changes the $d$ projected orbitals near the Fermi level, which further changes the MAE of bilayer $Cr_2Ge_2Te_6$ according to the following discussion.

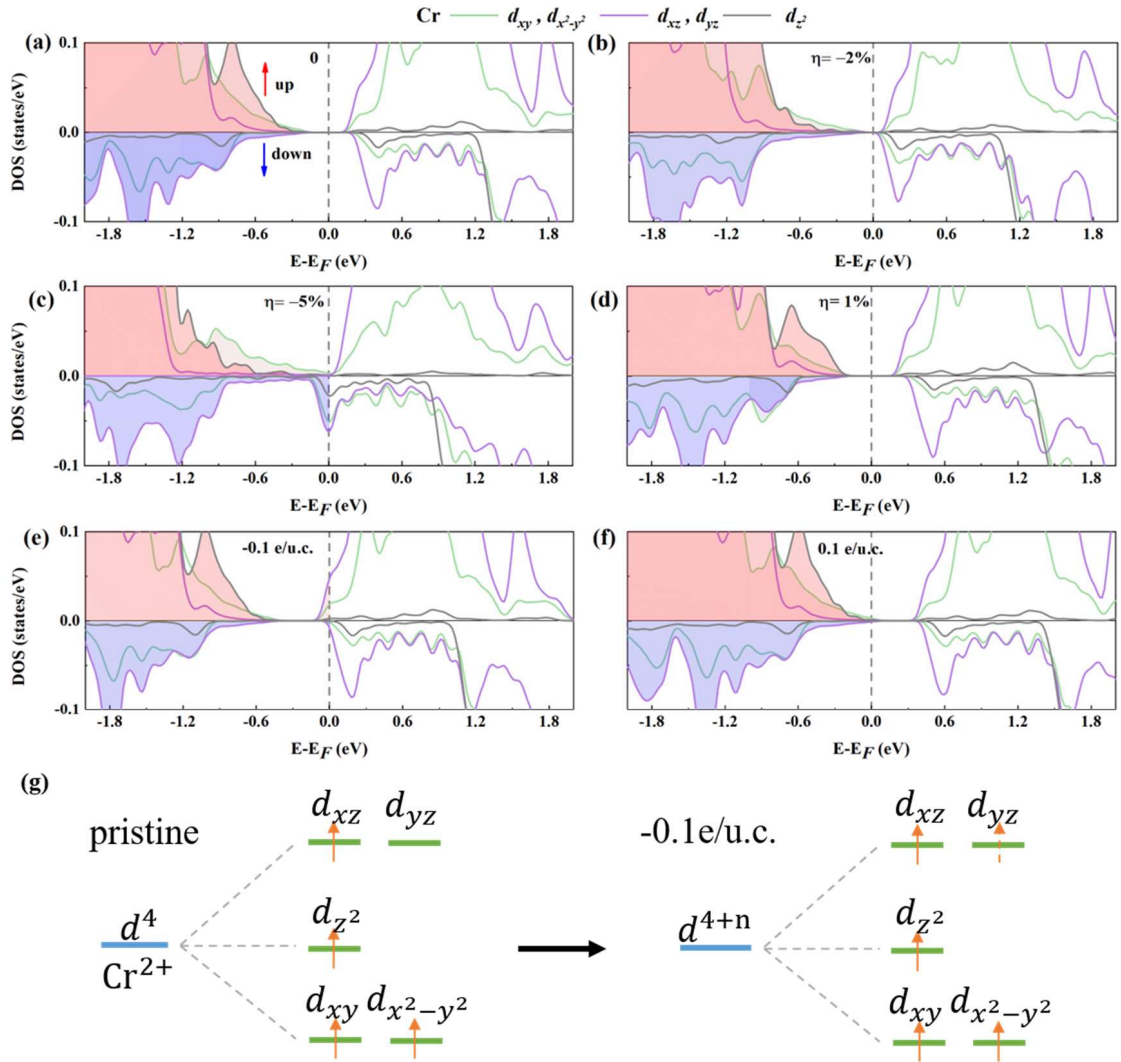

**Figure 6**. The spin-polarized projected density of states (PDOS) of $Cr_2Ge_2Te_6$-Cr for bilayer $Cr_2Ge_2Te_6$: (**a**) pristine, (**b**) compressive strain of 2%, (**c**) compressive strain of 5%, (**d**) tensile strain of 1%, (**e**) electron doping of 0.1 e/u.c, and (**f**) hole doping of 0.1 e/u.c. (**g**) The electronic configuration for pristine (left) and doping of -0.1 e/u.c. (right). The red and blue arrows indicate spin-up and spin-down states, respectively. The occupied states are presented by colorfully filled areas.

The MAE plays a crucial role in the stability of the long-range magnetic ground states [11,39]. We therefore investigated the variations of MAE and magnetic ground state with various strains or electrostatic doping concentrations in the bilayer $Cr_2Ge_2Te_6$. As shown in Fig. 7a, the ferromagnetic to antiferromagnetic transition can be induced by applying tensile strain more than 1%. Although larger $\alpha$ than 90° and longer Cr-Cr bond caused by the tensile strain indicating a weaker super-exchange interaction and a weaker direct interaction, combining the results in Fig. 3 and the ferromagnetic to antiferromagnetic transition shown in Fig. 7a, we can deduce that the direct exchange interaction dominates for tensile strain more than 1%. The ground state of the bilayer $Cr_2Ge_2Te_6$ is ferromagnetic when the applied strain is in the range of -4% to 1% while the magnetization direction remains out-of-plane. More meaningfully, we determine the critical strain range of -3% to 1%, within which MAE can be effectively enhanced comparing to the pristine one, and the ferromagnetic ground state maintains as well. From Fig. 7a, we can also see that large compressive strain (-5%) makes the direction of the easy axis change from out-of-plane to in-plane, which reduces the magnetic stability of the material. In the same way, we determined the critical doping range about 0 to 0.2 e/u.c, as shown in Fig. 7b. Beyond this range, the out-of-plane magnetization transformed into in-plane magnetization as electrostatic doping concentrations increase. Similarly, a range of electron doping from -0.25 e/u.c. to 0 is also predicted, within which the ferromagnetic stability can be maintained and the $T_c$ should be increased as the electron doping concentration increases due to the enlarged absolute value of $\Delta E_{FM-AFM}$ according to the mean field theory [40]. While the hole doping concentration has little effect on the magnetic ground state.

In order to see the insight of physical mechanisms beneath the variations of MAE, the second-order perturbation theory is engaged, which indicates that only the occupied and unoccupied Cr $d$ states near the Fermi level make major contributions to MAE in 2D magnetic systems [41]. Depending on the different spin channels, the contributions to MAE can be divided into two parts [41,42], including the same spin polarization and different spin polarizations, namely, $MAE = E_{\pm,\pm} + E_{\pm,\mp}$ [40], which are expressed by

$$E_{\pm,\pm} = (\xi)^2 \sum_{o^\pm, u^\pm} \frac{|<o^\pm|L_z|u^\pm>|^2 - |<o^\pm|L_x|u^\pm>|^2}{\varepsilon_u - \varepsilon_o}, \quad (1)$$

$$E_{\pm,\mp} = (\xi)^2 \sum_{o^\pm, u^\mp} \frac{|<o^\pm|L_x|u^\mp>|^2 - |<o^\pm|L_z|u^\mp>|^2}{\varepsilon_u - \varepsilon_o}, \quad (2)$$

where $o$ and $u$ denote the occupied and unoccupied states, respectively. The magnetization directions are denoted by $x$ and $z$. Positive sign represents spin-up states, and reversely, negative sign represents spin-down states. $\varepsilon_u$ ($\varepsilon_o$) stands for the energy of unoccupied (occupied) states, and the spin-orbit coupling constant is represented by $\xi$. Herein, five angular momentum matrix elements between two Cr $d$ orbitals are nonvanishing [41]: $<d_{xy}|L_x|d_{xz}>$, $<d_{z^2}|L_x|d_{yz}>$, $<d_{x^2-y^2}|L_x|d_{yz}>$, $<d_{xz}|L_z|d_{yz}>$, and $<d_{x^2-y^2}|L_z|d_{xy}>$. It is seen from Equation (1) that the out-of-plane spin polarization is favored for the occupied and unoccupied degenerate states due to the nonvanishing $<L_z>^2$ and vanishing $<L_x>^2$, while the in-plane spin polarization is favored with the nonvanishing $<L_x>^2$ for nondegenerate states. In contrary to Equation (1), for Equation (2), the out-of-plane spin polarization is favored for the occupied and unoccupied nondegenerate states due to the nonvanishing $<L_x>^2$ and vanishing $<L_z>^2$, while the in-plane spin polarization is favored with the vanishing $<L_x>^2$ and nonvanishing $<L_z>^2$ for degenerate states.

Based on above analyses, we can further understand the variations of MAE under different strains or electrostatic doping concentrations. For pristine bilayer $Cr_2Ge_2Te_6$, the spin-up Cr $d_{xy}/d_{x^2-y^2}$ orbitals contribute peaks in both VBM and CBM as shown in Fig. 6a, so the $<d_{x^2-y^2}|L_z|d_{xy}>$ in Equation (1) remained, leading to the positive MAE, which means the out-of-plane anisotropy is favored. With strain engineering ($\eta = -2\%$), we can see from the states around the Fermi level in Fig 6b that MAE is mainly contributed from $d_{yz}/d_{xz}/d_{xy}/d_{x^2-y^2}$ orbitals with different spin states, and the $<d_{xy}|L_x|d_{xz}>$ and $<d_{x^2-y^2}|L_x|d_{yz}>$ remained in Equation (2), so that we can obtain positive value of MAE, or in another words, the out-of-plane anisotropy is favored. Also, the reduced energy bandgap in Fig. 6b offers the decreased energy difference between the unoccupied and occupied states in Equation (2), which results in the enhanced ferromagnetic stability, identical with our results in Fig. 7a. With $\eta = -5\%$ shown in Fig. 6c, MAE are mainly derived from the inverse spin-polarized $d_{xz}/d_{yz}$, the $<d_{xz}|L_z|d_{yz}>$ in Equation (2) maintains, implying the in-plane anisotropy is favored, also identical with the results shown in Fig. 7a. From the results in Fig. 6d, the complex competition between orbital interactions may be the reason for maintaining out-of-plane anisotropy. At small electron doping concentrations (not larger than -0.1 e/u.c.), the Fermi level can only cross the spin-up states due to the obvious spin polarization in bilayer Cr$_2$Ge$_2$Te$_6$, the spin-up Cr $d_{yz}/d_{xz}/d_{xy}/d_{x^2-y^2}$ orbitals contribute peaks near the Fermi level in Fig. 6e, the out-of-plane anisotropy is favored because of the nonvanishing $<d_{xz}|L_z|d_{yz}>$ and $<d_{x^2-y^2}|L_z|d_{xy}>$ in Equation (2) [see Fig. 7b]. In Fig. 6f, the spin-up Cr $d_{xy}/d_{x^2-y^2}$ orbitals and spin-down Cr $d_{xz}/d_{yz}$ orbitals contribute peaks near the Fermi level, the $<d_{x^2-y^2}|L_z|d_{xy}>$ and $<d_{xz}|L_z|d_{yz}>$ in Equation (1) results in the positive MAE, *i.e.* the out-of-plane anisotropy [see Fig. 7b].

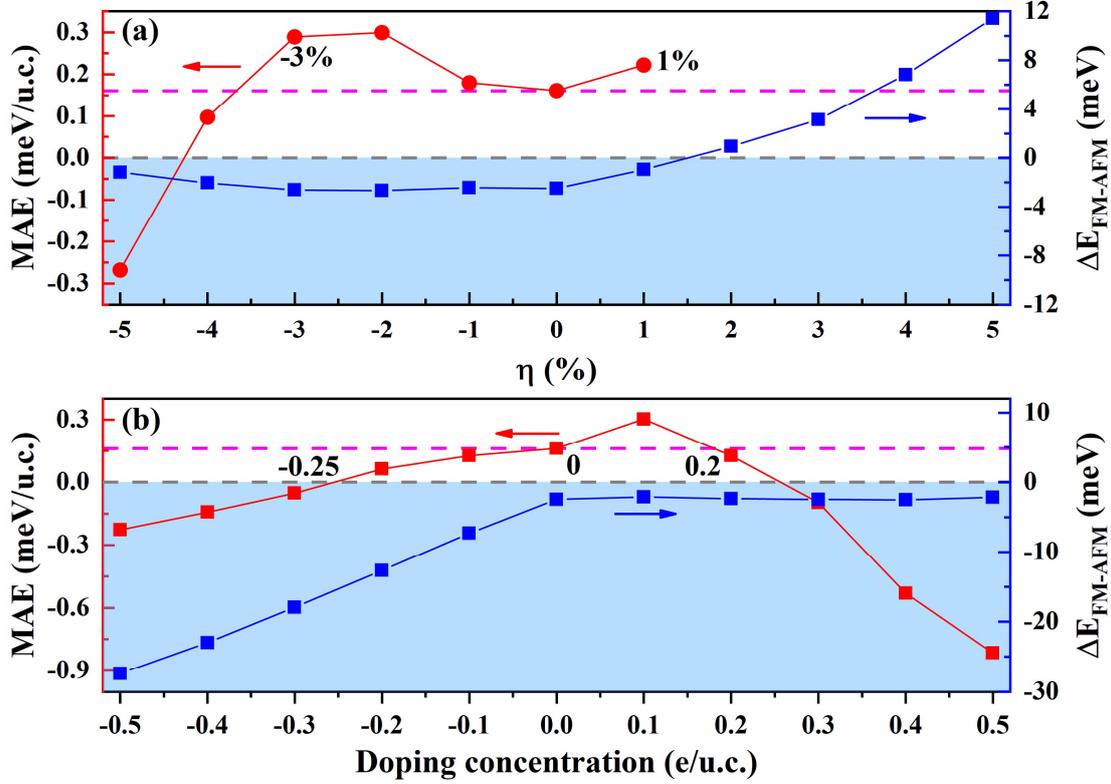

**Figure 7.** (**a**) Strains and (**b**) electrostatic doping concentration dependence of the magnetic anisotropy energy (MAE) and magnetic ground state for bilayer $Cr_2Ge_2Te_6$. The arrows point to the axis for each curve in the corresponding color. The gray horizontal dashed lines denote the boundary of spin polarization or magnetic ground state. The critical range is marked over the magenta dashed line in (**a**) and (**b**).

## Conclusion

In summary, we have studied the variations of electronic structures and magnetic properties of bilayer $Cr_2Ge_2Te_6$ with different strains or electrostatic doping concentrations. We proposed a critical strain ranges of -3% ~ 1% for bilayer $Cr_2Ge_2Te_6$, within which the ferromagnetic stability can be enhanced, as well as the critical doping range of 0 ~ 0.2 e/u.c. While beyond the critical range, the tensile strain induces a phase transition from the ferromagnetic to the antiferromagnetic, which is attributed to the competition between exchange interactions. Moreover, beyond the critical range, the compressive strain or electrostatic doping induced the magnetization direction to change from out-of-plane to in-plane. We also identified a range of electron doping from -0.25 e/u.c. to 0, within which the magnetic moment and $T_c$ can be increased, while the ferromagnetic stability was maintained. We have shown two ways for inducing half-metal in the bilayer $Cr_2Ge_2Te_6$. The compressive strain induced the reversed electron spin state at the conduction band minimum and the transition from semiconductor to metallic state. The second-order perturbation theory was applied to explain these variations of MAE. These results illustrated the tunability of electronic structures and magnetic properties by strain and electrostatic doping in the bilayer $Cr_2Ge_2Te_6$ and hopefully shed a light on achieving ferromagnetic stability for low-dimensional materials.

## Methods

Ab initio calculations were performed based on density functional theory. The exchange-correlation interaction was treated with the scheme of generalized gradient approximation (GGA) parametrized by the Perdew-Burke-Ernzerhof revised for solids (PBEsol) [43] as implemented in the Vienna ab initio Simulation Package (VASP) [44,45]. The accurate projector augmented wave method (PAW) [46] was employed for the following electronic configurations: $2p^63d^54s^1$ (Cr), $4s^24p^2$ (Ge), and $5s^25p^4$ (Te). A 500 eV kinetic energy cutoff of the plane-wave basis set was used for all calculations. The GGA+U was adopted for improving the description of on-site Coulomb interactions to the Cr $d$ orbital [47]. Different effective on-site Coulomb energy value $U_{eff}$ =U-J were conducted for magnetism, optimized lattice constants, and electronic structures with bilayer $Cr_2Ge_2Te_6$ (as shown in Supplementary Figs. S1, S2, and S3), which indicated that the results of $U_{eff}$ =1.7 eV were consistent with previous experiments and theoretical calculations [9,17,48]. The K-mesh of 7×7×1 was used for structural optimization and others were evaluated with a refined mesh of 20×20×1 subdivision in the full Brillouin zone. The maximum convergence force of all atoms was optimized until less than 0.01 eV/Å, and the convergence criterion for the energy differences was set as $1\times10^{-6}$ eV. To avoid the interaction between adjacent periodic layers, the vacuum space was included larger than 15 Å. To calculate the MAE, the spin-orbit coupling was taken into consideration. The crystal structures in this paper were drawn by VESTA package [49].

## References

1. Novoselov, K. S. *et al*. Electric feld efect in atomically thin carbon flms. *Science* **306**, 666–669 (2004).
2. Song, Y. H. *et al*. Observation of Coulomb gap in the quantum spin Hall candidate single-layer 1T'-WTe$_2$. *Nature Communications* **9**, 4071 (2018).
3. Deng, J. L. *et al*. Epitaxial growth of ultraflat stanene with topological band inversion. *Nature Materials*. **17**, 1081 (2018).
4. Li, L. *et al.* Black phosphorus feld-efect transistors. *Nature Nanotechnology*. **9**, 372 (2014).
5. Qiao, J., Kong, X., Hu, Z.-X., Yang, F. & Ji, W. High-mobility transport anisotropy and linear dichroism in few-layer black phosphorus. *Nature Communications* **5**, 4475, (2014).
6. Song C. S. *et al*. Tunable band gap and enhanced ferromagnetism by surface adsorption in monolayer $Cr_2Ge_2Te_6$. *Physical Review B* **99** 214435 (2019).
7. Sivadas N., Daniels M. W., Swendsen R. H., Okamoto S. & Xiao D., Magnetic ground state of semiconducting transition-metal trichalcogenide monolayers. *Physical Review B* **91**, 235425 (2015).
8. Li X. X. & Yang J. L., $CrXTe_3$ (X = Si, Ge) nanosheets: two dimensional intrinsic ferromagnetic semiconductors. *Journal of Materials Chemistry C* **2** 7071 (2014).
9. Fang Y. M., Wu S. Q., Zhu Z.-Z. & Guo G.-Y., Large magneto-optical effects and magnetic anisotropy energy in two-dimensional $Cr_2Ge_2Te_6$. *Physical Review B* **98** 125416 (2018).
10. Gong C. *et al*. Discovery of intrinsic ferromagnetism in two-dimensional van der Waals crystals. *Nature* (London) **546**, 265 (2017).
11. Mermin N. D. & Wagner H., ABSENCE OF FERROMAGNETISM OR ANTIFERROMAGNETISM IN ONE- OR TWO-DIMENSIONAL ISOTROPIC HEISENBERG MODELS. *Physical Review Letters* **17**, 1133 (1966).
12. Hatayama S. *et al*. Inverse Resistance Change $Cr_2Ge_2Te_6$-Based PCRAM Enabling Ultralow-Energy Amorphization. *ACS Applied Materials Interfaces* **10**, 2725 (2018).
13. Ji H. W. *et al*. A ferromagnetic insulating substrate for the epitaxial growth of topological insulators. *Journal of Applied Physics* **114**, 114907 (2013).
14. Liu Y. & Petrovic C., Anisotropic magnetic entropy change in $Cr_2X_2Te_6$ (X = Si and Ge). *Physical Review Materials* **3** 014001 (2019).
15. Pajda M., Kudrnovsky J., Turek I., Drchal V. & Bruno P., Oscillatory Curie Temperature of Two-Dimensional Ferromagnets. *Physical Review Letters* **85**, 5424-5427, (2000).
16. Irkhin V. Y., Katanin A. A. & Katsnelson M. I., Self-consistent spin-wave theory of layered Heisenberg magnets. *Physical Review B* **60**, 1082-1099 (1999).

17. Sun, Y. Y. *et al*. Electric manipulation of magnetism in bilayer van der Waals magnets. *J Phys Condens Matter* **31**, 205501, (2019).
18. Wang, Z. *et al*. Electric-field control of magnetism in a few-layered van der Waals ferromagnetic semiconductor. *Nat Nanotechnol* **13**, 554-559 (2018).
19. Xing, W. *et al*. Electric field effect in multilayer Cr2Ge2Te6: a ferromagnetic 2D material. *2D Materials* **4**, 024009 (2017).
20. Castro, E. V. *et al.* Biased bilayer graphene: Semiconductor with a gap tunable by the electric field effect. *Physical Review Letters* **99**, 216802 (2007).
21. Xia, F., Farmer, D. B., Lin, Y.-m. & Avouris, P. Graphene Field-Effect Transistors with High On/Off Current Ratio and Large Transport Band Gap at Room Temperature. *Nano Letters* **10**, 715-718, (2010).
22. Sun, Y. *et al*. Effects of hydrostatic pressure on spin-lattice coupling in two-dimensional ferromagnetic Cr2Ge2Te6. *Applied Physics Letters* **112**, 072409 (2018).
23. Lin, Z. *et al*. Pressure-induced spin reorientation transition in layered ferromagnetic insulator $Cr_2Ge_2Te_6$. *Physical Review Materials* **2**, 051004 (2018).
24. Ma, C., He, X. & Jin, K.-j. Polar instability under electrostatic doping in tetragonal SnTiO3. *Physical Review B* **96**, 035140 (2017).
25. Cao, T., Li, Z. & Louie, S. G. Tunable Magnetism and Half-Metallicity in Hole-Doped Monolayer GaSe. *Physical Review Letters* **114**, 236602, (2015).
26. Ma C., Jin K.-j., Ge C. & Yang G.-z. Strain-engineering stabilization of $BaTiO_3$-based polar metals. *Physical Review B* **97** 115103 (2018).
27. Wang, K. *et al*. Magnetic and electronic properties of Cr2Ge2Te6 monolayer by strain and electric-field engineering. *Applied Physics Letters* **114**, 092405 (2019).
28. Pustogow, A., McLeod, A. S., Saito, Y., Basov, D. N. & Dressel, M. Internal strain tunes electronic correlations on the nanoscale. *Sci. Adv.* **4**, 12, (2018).
29. Shukla, V., Grigoriev, A., Jena, N. K. & Ahuja, R. Strain controlled electronic and transport anisotropies in two-dimensional borophene sheets. *Physical Chemistry Chemical Physics* **20**, 22952-22960, (2018).
30. Fang, S., Carr, S., Cazalilla, M. A. & Kaxiras, E. Electronic structure theory of strained two-dimensional materials with hexagonal symmetry. *Physical Review B* **98**, 075106 (2018).
31. Jiang, L.-t. *et al*. Biaxial strain engineering of charge ordering and orbital ordering in HoNiO3. *Physical Review B* **97**, 195132 (2018).
32. Yang, D. *et al*. Cr2Ge2Te6: High Thermoelectric Performance from Layered Structure with High Symmetry. *Chemistry of Materials* **28**, 1611-1615, (2016).
33. Tian, Y., Gray, M. J., Ji, H., Cava, R. J. & Burch, K. S. Magneto-elastic coupling in a potential ferromagnetic 2D atomic crystal. *2D Materials* **3**, 025035 (2016).
34. Chen, X., Qi, J. & Shi, D. Strain-engineering of magnetic coupling in two-dimensional magnetic semiconductor CrSiTe3: Competition of direct exchange interaction and superexchange interaction. *Physics Letters A* **379**, 60-63, (2015).
35. Goodenough, J. B. Theory of the Role of Covalence in the Perovskite-Type

Manganites[La, M(II)]MnO$_3$. *Physical Review* **100**, 564-573, (1955)

36. J. B. Goodenough, AN INTERPRETATION OF THE MAGNETIC PROPERTIES OF THE PEROVSKITE-TYPE MIXED CRYSTALS La$_{1-x}$Sr$_x$CoO$_{3-\lambda}$ *J. Phys. Chem. Solids* **6**, 287 (1958).
37. J. Kanamori, SUPEREXCHANGE INTERACTION AND SYMMETRY PROPERTIES OF ELECTRON ORBITALS *J. Phys. Chem. Solids* **10**, 87 (1959).
38. Kang, S., Kang, S. & Yu, J. Effect of Coulomb Interactions on the Electronic and Magnetic Properties of Two-Dimensional CrSiTe$_3$ and CrGeTe$_3$ Materials. *Journal of Electronic Materials* **48**, 1441-1445, (2018).
39. Xu, C., Feng, J., Xiang, H. & Bellaiche, L. Interplay between Kitaev interaction and single ion anisotropy in ferromagnetic CrI$_3$ and CrGeTe$_3$ monolayers. *npj Computational Materials* **4**, 57 (2018).
40. Kittel C., *Introduction to Solid State Physics* (Wiley, New York, 2004).
41. Wang, D., Wu, R. & Freeman, A. J. First-principles theory of surface magnetocrystalline anisotropy and the diatomic-pair model. *Physical Review B Condens Matter* **47**, 14932-14947, (1993).
42. Lee, S.-C. *et al*. Effect of Fe–O distance on magnetocrystalline anisotropy energy at the Fe/MgO (001) interface. *Journal of Applied Physics* **113**, 023914 (2013).
43. Perdew, J. P. *et al*. Restoring the density-gradient expansion for exchange in solids and surfaces. *Physical Review Letters* **100**, 136406 (2008).
44. Kresse, G. & Furthmuller, J. Efficient iterative schemes for ab initio total-energy calculations using a plane-wave basis set. *Physical Review B* **54**, 11169-11186, (1996).
45. Kresse, G. & Furthmuller, J. Efficiency of ab-initio total energy calculations for metals and semiconductors using a plane-wave basis set. *Computational Materials Science* **6**, 15-50, (1996).
46. Kresse, G. & Joubert, D. From ultrasoft pseudopotentials to the projector augmented-wave method. *Physical Review B* **59**, 1758-1775, (1999).
47. Dudarev, S. L., Botton, G. A., Savrasov, S. Y., Humphreys, C. J. & Sutton, A. P. Electron-energy-loss spectra and the structural stability of nickel oxide: An LSDA+U study. *Physical Review B* **57**, 1505-1509, (1998).
48. Carteaux, V., Brunet, D., Ouvrard, G. & Andre, G. CRYSTALLOGRAPHIC, MAGNETIC AND ELECTRONIC-STRUCTURES OF A NEW LAYERED FERROMAGNETIC COMPOUND CR$_2$GE$_2$TE$_6$. *Journal of Physics-Condensed Matter* **7**, 69-87, (1995).
49. Momma, K. & Izumi, F. VESTA 3 for three-dimensional visualization of crystal, volumetric and morphology data. *Journal of Applied Crystallography* **44**, 1272-1276, (2011).

## Acknowledgments

This work was supported by the National Key Basic Research Program of China (Grant 2019YFA0308500), the National Natural Science Foundation of China (Grant

Nos. 11721404, 51761145104, 11974390 and 11674385), the Key Research Program of Frontier Sciences of the Chinese Academy of Sciences (Grant No. QYZDJ-SSW-SLH020), the Youth Innovation Promotion Association of CAS (Grant No. 2018008). We acknowledge Zhicheng Zhong and Peiheng Jiang for the helpful discussions.

## Author contributions

W.-N.R. and K.-J.J. contributed the whole idea and designed the research. W.-N.R., K.-J.J., and J.-S.W. wrote up the paper. W.-N.R. performed the theoretical calculations. C. M. prepared figure 5. K.-J.J. supervised the overall project. All authors discussed the results and commented on the manuscript.

## Competing interests

The authors declare no competing interests.

## Additional information

**Correspondence** and requests for materials should be addressed to K.-J.J.